\documentclass[letterpaper, 10 pt, conference]{ieeeconf}  % Comment this line out if you need a4paper

\IEEEoverridecommandlockouts                              % This command is only needed if 
\usepackage{graphicx}
\usepackage{amsmath}
\usepackage{xcolor}
\usepackage{ragged2e}
\usepackage{amssymb}
\usepackage{tabularx}
\usepackage{booktabs}
\usepackage{subfig}
\usepackage{empheq}
\usepackage{stackrel}
\usepackage[font={small,it}]{caption}
\usepackage{hyperref}
\usepackage{url}
\usepackage[most]{tcolorbox}

\newcommand{\myquote}[1]{``#1''}
\newcommand{\leftrarrows}{\mathrel{\raise.75ex\hbox{\oalign{%
  $\scriptstyle\leftarrow$\cr
  \vrule width0pt height.5ex$\hfil\scriptstyle\relbar$\cr}}}}
\newcommand{\lrightarrows}{\mathrel{\raise.75ex\hbox{\oalign{%
  $\scriptstyle\relbar$\hfil\cr
  $\scriptstyle\vrule width0pt height.5ex\smash\rightarrow$\cr}}}}
\newcommand{\Rrelbar}{\mathrel{\raise.75ex\hbox{\oalign{%
  $\scriptstyle\relbar$\cr
  \vrule width0pt height.5ex$\scriptstyle\relbar$}}}}

\newbox\tempbox
\newenvironment{nomenclature}{%
\newcommand\entry[2]{%
\setbox\tempbox\hbox{##1.\quad}
\hangindent\wd\tempbox\noindent{##1}\quad\ignorespaces##2\par}
\section*{Nomenclature}}{\par\addvspace{9pt}}

\newbox\tempbox
\newenvironment{notation}{%
\newcommand\entry[2]{%
\setbox\tempbox\hbox{##1.\quad}
\hangindent\wd\tempbox\noindent{##1}\quad\ignorespaces##2\par}
\section*{Notation}}{\par\addvspace{9pt}}

\newtcolorbox{mylunchbox}[1][]{%
    colback=black!12,
    colframe=black!12,
    notitle,
    borderline west={0pt}{0pt}{black!12},
    enhanced,
    breakable,
    left=0pt,
right=0pt,
top=0pt,
bottom=0pt
    }

\title{\LARGE \bf Combining physics-based and machine learning methods to accelerate innovation in sustainable transportation and beyond: a control perspective}

\author{Gabriele Pozzato$^1$ and Simona Onori$^{1,*}, IEEE\,\,Senior\,Member$
\thanks{$^{1}$ Energy Science \& Engineering, Stanford University, Stanford, CA.}
\thanks{$^{*}$ corresponding author  {\tt\small sonori@stanford.edu}}}

\begin{document}
\maketitle
\thispagestyle{empty}
\pagestyle{empty}
%%%%%%%%%%%%%%%%%%%%%%%%%%%%%%%%%%%%%%%%%%%%%%%%%%%%%%%%%%%%%%%%%%%%%%%%%%%%%%%%
\begin{abstract}
Lithium-ion batteries are playing a key role in the sustainable energy transition.  To fully exploit the potential of this technology, a variety of modeling,  estimation,  and prediction problems need to be addressed to enhance its design and optimize its utilization.  
Batteries are complex electrochemical systems whose behavior drastically changes as a function of aging,  temperature,  C-rate,  and state of charge,  posing unique modeling and control research questions.

In this tutorial paper,  we provide insights into three battery modeling methodologies,  namely first principle,  machine learning,  and hybrid modeling. Each approach has its own strengths and weaknesses, and by means of three case studies we describe main characteristics and challenges of each of the three methods.
\end{abstract}

%%%%%%%%%%%%%%%%%%%%%%%%%%%%%%%%%%%%%%%%%%%%%%%%%%%%%%%%%%%%%%%%%%%%%%%%%%%%%%%%
\section{Introduction and Motivation}
According to a recent McKinsey\&Company \cite{newmckinsey} report,  the lithium-ion battery chain,  from mining through recycling,  is projected to have an annual growth of over 30\% from 2022 to 2030 and reach a market size of 4.7TWh.  To support the transition to electrified transportation systems and {decarbonization},  high specific power and energy storage devices are needed.  Compared to other electrochemical solutions available on the market,  lithium-ion batteries {(LIBs) -- with power and energy density} of  300-1500W/kg and 100-250Wh/kg \cite{xia2019practical},  respectively -- play a dominant role.

{A LIB is composed of positive and negative electrodes,  a separator, an electrolyte, and current collectors. }  In a cell,  lithium ions are {shuffled} between the electrodes through the separator,  a  permeable membrane placed between the electrodes,  {while} electrons flow through an external circuit.  Electrochemical models are well-established approaches to describe lithium transport in the liquid phase, and intercalation in the negative and positive electrodes (the solid phase).  Electrochemical models based on the conservation of mass and charge {partial differential equations (PDEs)} are in the form of single-particle (SPM) \cite{haran1998determination,santhanagopalan2006review},  enhanced single particle (ESPM) \cite{tanim2015temperature},  and Doyle-Fuller-Newman (DFN) \cite{doyle1993modeling} models. These models are effective tools to describe the underlying electrochemical phenomena {under} different operating conditions and can be extended to account for aging modes.  
%In \cite{arora1999mathematical},  lithium plating and solid electrolyte interface (SEI) growth were introduced in the DFN model to describe the battery aging.  Similarly,  \cite{pannala2022methodology} models SEI and loss of active material (LAM) relying on a SPM.  A comprehensive model coupling SEI formation,  lithium plating,  and loss of active material in the ESPM framework is developed in \cite{pozzato2021modeling}.  
The key limitation of electrochemical models is overparametrization and parameters identifiability \cite{schmidt2010experiment},  which could hinder their use in particular when cell's properties are changing over time.

In the last years, machine learning models  have gained {attention} as promising tools for battery state of health (SOH) estimation and residual useful life (RUL)  prediction \cite{sulzer2021challenge}.  Provided that a comprehensive dataset is available,  this modeling approach is appealing because does not require a deep understanding of electrochemical kinetics and transport phenomena {  and is characterized by a relatively short development time.  The model complexity is not fixed and can be chosen through optimization routines such as grid search.} Moreover,  these models do not require the solution of PDEs and,  if trained properly,  can achieve high accuracy with low computational cost.  {Machine learning models are black box models,}  i.e.,  they use historical data to learn the input-output behavior of the system and do not provide any insight into the underlying physics.  A key limitation of machine learning approaches is {their limited or nonexistent extrapolation capability} \cite{reyes2019machine}.  The most common solution to this problem is to increase the cardinality of the training dataset in order to cover the whole operating region of the battery (e.g.,  in terms of current,  voltage,  {state of charge (SOC), } and temperature) \cite{song2020intelligent}.  However,  this is not always a viable option,  as it could lead to long and costly experimental campaigns \cite{ngspectrum}.  Another solution is to carefully engineer experiments to collect data carrying information {on the phenomena of interest.}
%This \myquote{data-centric} approach reduces the testing time while creating lower dimensional and highly informative datasets.  

%\begin{figure*}[!t]
%\centering 
%\includegraphics[width=\textwidth]{figure/framework.pdf}
%\caption{\textbf{Physics-based, machine learning,  and hybrid modeling. } (a) In physics-based models,  complexity is usually fixed and models are selected beforehand.  After that,  parameters are identified with procedures such as the least squares method or evolutionary algorithms.  (b) In machine learning models,  only the model class is selected (i.e.,  neural networks) and optimization strategies -- like grid search -- are  used to optimize the hyperparameters of the model and its complexity.  (c) Hybrid models blend the strengths of physics-based and machine learning approaches.}
%\label{fig:framework}	
%\end{figure*}

To blend the insight of physics-based models with the learning {capability of machine learning tools},  hybrid models have recently been proposed to innovate the field of battery modeling and life prediction.   Generally speaking,  hybrid modeling frameworks have the potential to increase the predictive and extrapolation capabilities of battery models.  {However,  a range of different configurations combining the strengths of the two methods can be used based upon the specific application \cite{aykol2021perspective}.}

Batteries are complex electrochemical systems affected by different aging modes and their behavior drastically changes as a function of temperature,  C-rate,  {SOC,  and time}.  We believe that problems related to electrochemical modeling, estimation, or prediction do not have a single all-encompassing solution.  Following this rationale,  there is no clear winner between physics-based, machine learning,  and hybrid modeling;  instead,  these approaches complements each other and should be carefully exploited depending on the problem to be solved.  

In this tutorial,  we dig further into each {approach} and show three case studies exemplifying the application of each strategy.  In Section \ref{cs1},  an electrochemical model for NMC batteries integrating {different} degradation modes is shown.   In Section \ref{cs2},  offline and online battery health estimation strategies for second-life applications are considered.  Lastly,  in Section \ref{cs3},  a hybrid solution blending physics with machine learning for the problem of LFP batteries modeling is described.  
%In this context,  an electrochemical model based on PDEs is used to describe mass and charge transport within the battery.  While preserving the physical understanding of electrochemical models, this framework exploits the flexibility of machine learning to describe the battery voltage hysteresis.

\section{Physics-based modeling}\label{sec:pb}
\vspace{0.5em}
\begin{mylunchbox}
\textbf{Pros:}
\begin{itemize}
\item Physical insight
\item Good accuracy over wide operating regions
\item Can include degradation mechanisms
\item {Extrapolation capability}
\item {Transferability across chemistries}
\end{itemize}
\end{mylunchbox}

\begin{mylunchbox}
\textbf{Cons:}
\begin{itemize}
\item High computational {cost}
\item Overparametrization and parameter identifiability
\item {Limited understanding of some degradation modes}
\end{itemize}
\end{mylunchbox}

%\textbf{Talk about PB in general and literature review}
Electrochemical models are used to predict lithium transport in the electrolyte phase and intercalation/deintercalation in the solid phase.  The solid phase is composed of  particles with different shapes and sizes packed together to form a porous structure which is saturated with electrolyte (the liquid phase).  {Particles in the electrodes are usually modeled as spheres and,  depending on the number of particles,  models can be divided in two groups: the ones considering an ensemble of particles (e.g.,  DFN \cite{doyle1993modeling,torchio2016lionsimba}),  and the ones lumping the electrode in one single particle (e.g.,  SPM and ESPM \cite{haran1998determination,santhanagopalan2006review,tanim2015temperature}). }  Depending on the selected modeling strategy,  different levels of accuracy in the description of electrochemical dynamics can be achieved.

For the purpose of this tutorial paper,  we focus on the use of ESPM to solve problems related to the modeling of aging modes in Li-ion batteries.  {ESPM is a trade-off} between the complexity of the multi-particle DFN model and the simplicity of SPM (which neglects charge conservation in solid and electrolyte phases,  and mass conservation in the electrolyte phase).  Specifically,  in the ESPM mass balance equations are used to describe lithium concentration in both solid and electrolyte phases, and conservation of charge is modeled in the electrolyte phase only.  The description of other modeling strategies is out of the scope of this work and interested readers are referred to \cite{lee2022advanced}. 

\begin{figure}[!tb]
\centering 
\includegraphics[width=0.85\columnwidth]{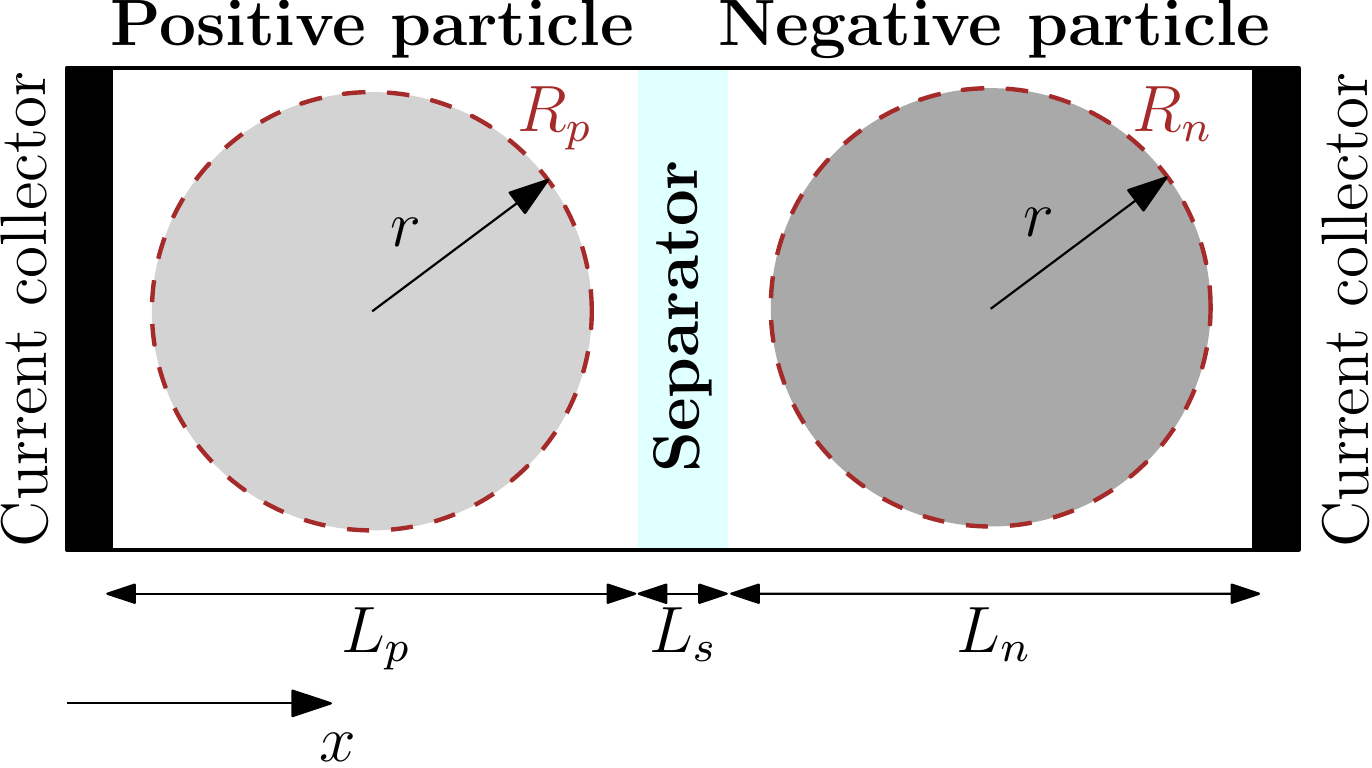}
\caption{{\textbf{Battery schematic for ESPM. } In the ESPM,  positive and negative electrodes are modeled as single particles divided by a separator and immersed in the electrolyte. }}
\label{fig:espm}	
\end{figure}

In the ESPM framework,  positive and negative electrodes are modeled as single particles,  as shown in Figure \ref{fig:espm}.  During charge,  lithium deintercalates the positive electrode,  is transported in the electrolyte (in a solvation shell),  and then intercalates the negative electrode.  During discharge,  the opposite process takes place and lithium is deintercalated from the negative electrode and intercalated in the positive one.  The whole process is described by means of mass and charge balance equations.   \\

\noindent\textbf{Mass balance in the solid phase \cite{rahn2013battery}.} In the solid phase,  mass balance equations model the lithium intercalation in the host material by means of the Fick's law of diffusion (in spherical coordinates):
\begin{equation}
\frac{\partial c_{s,i}}{\partial t} = \frac{1}{r^2}\frac{\partial}{\partial r}\left(r^2D_{s,i}\frac{\partial c_{s,i}}{\partial r}\right)
\label{eq:mb1}
\end{equation}
where $i\in \{p,n\}$ ($p$: positive electrode,  $n$: negative electrode),  $R_{i}$ is the particle radius,  $r\in (0, R_{i})$ is the radial coordinate,  $c_{s,i}$ is the concentration of lithium in the particle,  and $D_{s,i}$ is the solid-phase diffusion coefficient.  At the particle center ($r = 0$) and surface ($r = R_{i}$),  the following boundary conditions are enforced:
\begin{equation}
\begin{split}
&\textit{Zero-flux:}\\
&\quad \frac{\partial c_{s,i}}{\partial r}\bigg|_{r=0} = 0\\
&\textit{Flux inside/outside the particle:}\\
&\quad \frac{\partial c_{s,i}}{\partial r}\bigg|_{r=R_{i}} =  f(I)\hspace{13em}
\end{split}
\end{equation}
where $I$ is the applied current and $f$ is a function modeling the flux of lithium going inside and outside the particle.  In the remainder of the paper,  slightly different expressions for $f(I)$ are used depending on whether the battery is experiencing aging or not.\\

\noindent\textbf{Mass balance in the electrolyte phase \cite{rahn2013battery}.} The mass balance equation in the electrolyte phase is formulated as follows:
\begin{equation}
\varepsilon_i\frac{\partial c}{\partial t} = -\frac{\partial}{\partial x}\left(-D_{eff,i}\frac{\partial c}{\partial x}\right) + (1-t_+)J_i
\label{eq:mb2}
\end{equation}
with $i\in\{p,s,n\}$ ($s$: separator),  $x$ the Cartesian coordinate defined inside $(0,L_p)$,  $(L_p,L_p+L_s)$,  and $(L_p+L_s,L_p+L_s+L_n)$ for positive electrode,  separator, and negative electrode,  respectively,  $\varepsilon_i$ the porosity of the electrode (i.e.,  the space occupied by the electrolyte),  $J_i$ the flux of ions from/to the solid phase,  and $t_+$ the transference number (modeling the lithium transferred by migration rather than diffusion).  The term $-D_{eff,i}\frac{\partial c}{\partial x}$ models the diffusion flux caused by the concentration gradient within the electrolyte,  i.e.,  ions are forced to move to regions where concentration is lower.  This motion is a function of the effective diffusion coefficient ($D_{eff,i}$) computed as follows:
\begin{equation}
D_{eff,i} = \frac{\varepsilon_i}{\tau_i}D_i= \varepsilon_i^{brugg}D_i
\label{eq:diffeff}
\end{equation}
with $D_i$ the bulk electrolyte diffusivity,  $\tau_i$ the tortuosity,  and $brugg$ the Bruggeman coefficient,  equal to 1.5 for porous structures composed of uniform-sized spherical particles \cite{suthar2015effect}.  At the current collectors ($x=0$ and $x = L_p+L_s+L_n$) and interfaces,  the following boundary conditions are enforced: 
\begin{equation}
\begin{split}
&\textit{Zero-flux:}\\
&\quad\quad \frac{\partial c}{\partial x}\bigg|_{x=0} = 0, \quad \frac{\partial c}{\partial x}\bigg|_{x=L_p+L_s+L_n} = 0\\ 
&\textit{Continuity (concentration):}\\
&\quad\quad c|_{x=L_p^-} = c|_{x=L_p^+}, \quad c|_{x=L_p+L_s^-} = c|_{x=L_p+L_s^+}\\
&\textit{Continuity (flux):}\\
&\quad\quad -D_{eff,p}\frac{\partial c}{\partial x}\bigg|_{x=L_p^-} = -D_{eff,s}\frac{\partial c}{\partial x}\bigg|_{x=L_p^+},\\
&\quad\quad -D_{eff,s}\frac{\partial c}{\partial x}\bigg|_{x=L_p+L_s^-} = -D_{eff,n}\frac{\partial c}{\partial x}\bigg|_{x=L_p+L_s^+}\hspace{0.1em}
\end{split}
\end{equation}\\

\noindent\textbf{Conservation of charge in the electrolyte phase \cite{rahn2013battery}.} Electrodes are porous media where electrolyte flows in the pores allowing charge carriers (lithium ions) to diffuse.  Hence,  concentration of charge must account for both electrostatic and diffusional effects and takes then following expression:
\begin{equation}
  \resizebox{0.9\columnwidth}{!}{$
\kappa_{eff,i}\frac{\partial}{\partial x}\left(\frac{\partial \phi_e}{\partial x}\right) - \frac{2RT\kappa_{eff,i}v}{F}\frac{\partial^2\ln(c)}{\partial x^2}=-FJ_i$}
\label{eq:cb1}
\end{equation}
where $\phi_e$ is the electrolyte potential,  $R$ is the universal gas constant,  $T$ is the temperature,  $F$ is the Faraday constant,  $J_i$ is the flux of ions from/to the solid phase,  $v$ is the thermodynamic factor,  and $\kappa_{eff,i} = \varepsilon_i^{brugg}\kappa$ is the effective electrolyte conductivity defined as a function of the bulk electrolyte conductivity $\kappa$ (similarly to Equation \eqref{eq:diffeff}). \\
The first term on the left-hand side accounts for electrostatics,  i.e.,  for the voltage variation associated with the electrolyte conductivity.  The second term on the left-hand side models the interplay between the concentration gradient of lithium ions,  the electric field,  and the potential.  Eventually,  the term on the right-hand side models the flux of current from the solid phase.  At the current collectors ($x=0$ and $x = L_p+L_s+L_n$) and interfaces,  the following boundary conditions are enforced: 
\begin{equation}
\begin{split}
&\textit{Zero-flux:}\\
&\quad\quad \frac{\partial \phi_e}{\partial x}\bigg|_{x=0} = 0, \quad \frac{\partial \phi_e}{\partial x}\bigg|_{x=L_p+L_s+L_n} = 0\\ 
&\textit{Continuity (potential):}\\
&\quad\quad \phi_e|_{x=L_p^-} = \phi_e|_{x=L_p^+}, \quad \phi_e|_{x=L_p+L_s^-} = \phi_e|_{x=L_p+L_s^+}\\
&\textit{Continuity (flux):}\\
&\quad\quad -\kappa_{eff,p}\frac{\partial  \phi_e}{\partial x}\bigg|_{x=L_p^-} = -\kappa_{eff,s}\frac{\partial  \phi_e}{\partial x}\bigg|_{x=L_p^+},\\
&\quad\quad -\kappa_{eff,s}\frac{\partial  \phi_e}{\partial x}\bigg|_{x=L_p+L_s^-} = -\kappa_{eff,n}\frac{\partial  \phi_e}{\partial x}\bigg|_{x=L_p+L_s^+} 
\end{split}
\end{equation}\\

\noindent\textbf{Butler-Volmer kinetics \cite{rahn2013battery}.}  During intercalation/deintercalation reactions,  electrons must cross the electrode-electrolyte interface \cite{rahn2013battery}.  The Butler-Volmer kinetics provides the link between overpotential and intercalation/deintercalation currents and takes the following expression:
\begin{equation}
\mathcal{J}_i = i_{0,i} \left[\exp\left(\frac{\alpha_a F}{RT}\eta_i\right)-\exp\left(\frac{-\alpha_cF}{RT}\eta_i\right)\right]
\label{bv}
\end{equation}
where $i_{0,i}$ is the exchange current density,  $\eta_i$ is the overpotential,  and $\alpha_a$ and $\alpha_c$ are the anodic and cathodic charge transfer coefficients.   Assuming $\alpha_a=\alpha_c = 0.5$ \cite{prada2012simplified},  Equation \eqref{bv} can be rewritten as follows:
\begin{equation}
\resizebox{\columnwidth}{!}{$
\eta_i = \frac{RT}{0.5F}\sinh^{-1}\left(\frac{\mathcal{J}_i}{2i_{0,i}}\right) = \frac{RT}{0.5F}\sinh^{-1}\left(\frac{\mp I}{2A_{cell}a_iL_ii_{0,i}}\right)
$}
\end{equation}
with $A_{cell}$ the cell cross section area,  $a_i$ the specific surface area,  and $\mathcal{J}_i$ defined as $\frac{\mp I}{A_{cell}a_{i}L_i}$\footnote{In Section \ref{cs1},  $a_{i,t}$ is used in place of $a_{i}$ in the denominator of $\mathcal{J}_i$.}.\\

One key challenge of electrochemical models is parameter identification.  Generally,  electrochemical models are overparameterized,  which can lead to a lack of identifiability  {especially when only current and voltage data are available.  When the model's parameters can not be uniquely identified from data,  the low information content from data or the weak structural model properties cause the lack of identifiability \cite{ljung1987theory}.  } The literature on this topic is broad and providing a complete picture is out of the scope of this tutorial paper.  {Typical approaches to assess parameters' identifiability in electrochemical models involve the use of Fisher information \cite{schmidt2010experiment,forman2012genetic, song2018parameter},  sensitivity matrix \cite{lai2020analytical},  and correlation analysis of the sensitivity matrix \cite{allam2020online}.  
To identify the model parameters,  a nonlinear optimization problem is formulated and solved relying on methods such as evolutionary algorithms \cite{zhang2014multi, pozzato2022core},  Gauss-Newton method \cite{ramadesigan2011parameter},  sequential quadratic programming \cite{lee2020electrode},  and trust region reflective algorithm \cite{marcicki2012nonlinear}.  More advanced techniques based on pulse testing and electrochemical impedance spectroscopy (EIS) -- to estimate a subset of the model's parameters --  can be found in \cite{lu2021nondestructive} and \cite{lu2022nondestructive},  respectively.}

\subsection{Case study: NMC electrochemical and aging model}\label{cs1}
% Rewrite (nomenclature? coerente con hybrid model? Spiegare tutto perchè è un tutorial)
% Spiegare parte relativa a aging
Equations \eqref{eq:mb1},  \eqref{eq:mb2},  and \eqref{eq:cb1} are the building blocks to develop fresh battery cell models.  However,  during operation,  batteries experience performance  degradation in the form of capacity and power fade.  {The root causes of this degradation are a variety of aging modes which can be  grouped into two classes:} loss of lithium inventory (LLI) and loss of active material (LAM) \cite{birkl2017degradation}.  LLI is associated with the loss of cyclable lithium during battery operation,  whereas during LAM active material becomes unavailable for lithium insertion. For an effective management and operation of LIB,  it is crucial to understand and model the underlying aging modes leading to LLI and LAM.  

%{This could allow for the development of state of health estimation strategies decoupling the contribution of each aging modes,  leading to a safer battery operation.  }

\begin{figure}[!tb]
\centering 
\includegraphics[width=\columnwidth]{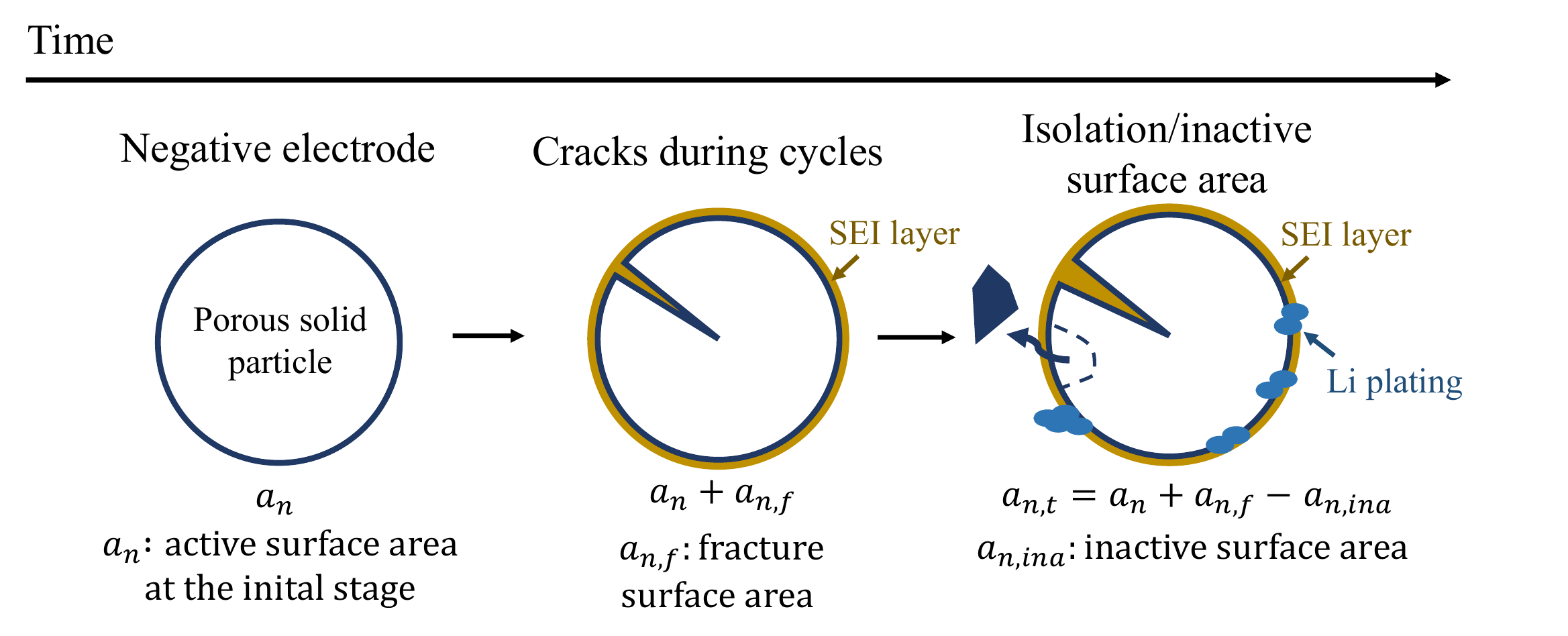}
\caption{\textbf{Aging modes in the negative electrode \cite{pozzato2021modeling}. }  On the left, the fresh particle is shown.  At the early stage of life,  the particle experiences SEI formation and cracks induced by mechanical stress (in the middle).  After prolonged cycling,  the particle shows increased SEI growth and lithium plating (on the right).}
\label{fig:expmnmc}	
\end{figure}

In \cite{pozzato2021modeling},  we formulate a physics-based modeling framework that integrates {ESPM with} SEI layer growth,  lithium plating,  and particle fracture/isolation.   Lithium plating and SEI dynamics are associated with LLI in the negative electrode.  SEI is a passivation layer formed at the surface of the negative electrode's particles 
and created by the spontaneous reaction of the solvent (a component of the electrolyte) with the active material.  Lithium plating is defined as the formation of metallic lithium on the negative electrode of LIBs during charging.  This phenomenon not only causes capacity and power fade but also poses  significant safety concerns when unmonitored: plating can lead to irregular dendrite growth that can pierce the separator causing thermal runaway and catastrophic failure of the battery cell.  Particle fracture and isolation are two competing mechanisms leading to increase and decrease of the active electrode surface area,  respectively.  The increase in surface area created during fracture leads to additional SEI formation which contributes to LLI.  {Electronic isolation leads to portions of the electrode becoming electrically isolated and,  therefore,  not accessible by lithium ions \cite{narayanrao2012phenomenological}. } This leads to LAM but also to the modification of the active surface area,  which leads to a modification of porosity,  SEI layer growth dynamics,  and lithium plating.   While SEI formation and plating are modeled on the negative electrode only,  fracture and isolation happen on both electrodes.  In \cite{pozzato2021modeling},  we focused on the analysis of the negative electrode's aging modes,  which are shown graphically in Figure \ref{fig:expmnmc}.

Mathematically,  SEI and lithium plating are described by Equations \eqref{eq:SEI_1} to \eqref{eq:SEI_3}.  Equation \eqref{eq:SEI_1} is based on the Tafel law  and models the side current associated with SEI formation.  Similarly,  the side current due to irreversible lithium plating is modeled by the Tafel law in Equation \eqref{eq:lpl_1}.  Concentrations of SEI and plated lithium are described by Equations \eqref{eq:SEI_2}  and \eqref{eq:lpl_2}.  If the plated lithium reacts with the electrolyte,  plating contributes to the formation of the SEI layer with the term $\frac{j_{lpl}}{2F}\beta_{lpl}$ in Equation \eqref{eq:SEI_2}.  SEI and plated lithium form a film on the surface of the particle which,  according to \cite{yang2017modeling},  is modeled by Equation \eqref{eq:SEI_3}.  In this film,  the resistance associated with SEI is modeled with Equation \eqref{eq:SEI_4}.
Isolation and fracture are described using the model proposed by \cite{narayanrao2012phenomenological} and shown in Equation \eqref{eq:lam_1}.  \\
Mass and charge balance equations for the ESPM (developed from Equations \eqref{eq:mb1},  \eqref{eq:mb2},  and \eqref{eq:cb1}),  battery terminal voltage,  and SOC are shown Table \ref{table:ESPM_table_1}.  Models of degradation mechanisms are collected in Table \ref{table:ESPM_table_2}.  

\begin{table}[!tb]
	\caption{ESPM governing equations.}\label{table:ESPM_table_1}
	\Centering	
	\resizebox{\columnwidth}{!}{
	\scriptsize{				
		\begin{tabular}{l}
			\hline\hline \\ [-3mm]
           \parbox{8cm}{
			    \begin{equation}
			    \label{goveq:eq1}
			    \begin{split}
			    &J_n = \frac{I}{A_{cell}FL_n},\quad J_p = \frac{-I}{A_{cell}FL_p}, \quad J_s = 0\hspace{7em}\\
			    \end{split}
				\end{equation}}\\
             \textbf{Mass transport in the electrolyte phase, } $i\in \{ p,s,n\}$ \\
			\parbox{8cm}{
			    \begin{equation}
			     \label{goveq:eq2}
			    \begin{split}
			    & \varepsilon_i\frac{\partial c}{\partial t} = \frac{\partial}{\partial x}\left(D_{eff,i}(c,T)\frac{\partial c}{\partial x}\right) + (1-t_+)J_i\hspace{9.5em}\\
			    \end{split}
				\end{equation}}\\
		    \textbf{Charge transport in the electrolyte phase, } $i\in\{p,s,n\}$ \\ 
		    \parbox{8cm}{
			    \begin{equation}
			     \label{goveq:eq3}
			      \resizebox{0.9\columnwidth}{!}{$
			    \begin{split}
			    &\kappa_{eff,i}(c^{avg}_i,T)\frac{\partial}{\partial x}\left(\frac{\partial \phi_e}{\partial x}\right) - \frac{2\kappa_{eff,i}(c^{avg}_i,T)RT}{F}(1-t_+)\frac{\partial^2\ln c}{\partial x^2}+FJ_i=0\\
			    \end{split}$}
				\end{equation}} \\
		    \textbf{Mass transport in the solid phase,} $i\in\{p,n\}$ \\ 
		    \parbox{8cm}{
			    \begin{equation}
			    \label{goveq:eq4} 
			    \frac{\partial c_{s,i}}{\partial t} = \frac{1}{r^2}\frac{\partial}{\partial r}\left(r^2D_{s,i}(T)\frac{\partial c_{s,i}}{\partial r}\right);\quad  \frac{\partial c_{s,i}}{\partial r}\bigg |_{r=0}=0 \hspace{5em}
				\end{equation}}\\ 
		     \parbox{8cm}{
			    \begin{equation}
			    \label{goveq:eq5} 
			    \begin{split}
			    \frac{\partial c_{s,p}}{\partial r}\bigg\vert_{r = R_p} \hspace{-1.5em}= \frac{I}{D_{s,p}(T)a_{p,t}A_{cell}FL_p}\hspace{12em}\\  
			    \frac{\partial c_{s,n}}{\partial r}\bigg\vert_{r = R_n} \hspace{-1.5em}= \frac{-I+L_nA_{cell}(j_{SEI}+j_{lpl})}{D_{s,n}(T)a_{n,t}A_{cell}FL_n} \hspace{9.2em}
			    \end{split}
				\end{equation}}\\
				\textbf{Terminal voltage} \\ 
			\parbox{8cm}{
			    \begin{equation}
					V = U_p - U_n + \eta_p - \eta_n + \Delta\Phi_e - I\cdot (R_l+R_{el}+R_{film})
				\end{equation}}\\
				\textbf{State of charge} \\ 
			\parbox{8cm}{
			    \begin{equation}
			    \label{goveq:eqsoc} 
					SOC_n = \frac{\theta_n-\theta_{n,0\%}}{\theta_{n,100\%}-\theta_{n,0\%}},\  SOC_p = \frac{\theta_{p,0\%}-\theta_p}{\theta_{p,0\%}-\theta_{p,100\%}}\hspace{2em}
				\end{equation}}\\ [3mm]
				
			\hline\hline 
		\end{tabular}}}
		\vspace{-2em}
\end{table}

\begin{table}[t]
	\caption{ESPM degradation modes.}\label{table:ESPM_table_2}
	\centering
		\renewcommand{\arraystretch}{1.4}
		\resizebox{0.49\textwidth}{!}{		
		\begin{tabular}{l}
			\hline\hline\\
			\multicolumn{1}{l}{\textbf{Porosity variation}}  \\ [-2mm]
			\parbox{9.5cm}{
			    \begin{flalign} \label{eq:poro}
				& \quad \varepsilon_{p} = \varepsilon_{p,0}+\frac{a_{p,ina}-a_{p,f}}{3}R_p,\\
				& \quad \varepsilon_{n} = \varepsilon_{n,0}+\frac{a_{n,ina}-a_{n,f}}{3}R_n - \nu_n\frac{3L_{film}}{R_n}&
				\end{flalign}} \\[-2mm]
			\multicolumn{1}{l}{\textbf{SEI layer growth}}  \\ [-2mm]
			\parbox{9.5cm}{
			    \begin{flalign} \label{eq:SEI_1}
				&  j_{SEI} = -F\ a_{n,t}\ k_f(c_{s,n},T)\ c_{solv}^{surf}\exp\left[-\frac{\alpha_s F}{RT}\left(\Phi_{s,n}-R_{film}I\right)\right]
				\end{flalign}} \\[-2mm]
			\parbox{9.5cm}{
			    \begin{flalign} \label{eq:SEI_2}
				& \quad \frac{dc_{SEI}}{dt} = -\left(\frac{j_{SEI}}{2F}+\frac{j_{lpl}}{2F}\beta_{lpl}\right) &
				\end{flalign}} \\[-2mm]
		    \parbox{9.5cm}{
			    \begin{flalign} \label{eq:SEI_4}
				& \quad R_{film} = \frac{L_{SEI}}{a_{n,t}\ A_{cell}\ L_n\ \kappa_{SEI}} &
				\end{flalign}} \\[-2mm]
				
			\multicolumn{1}{l}{\textbf{Li plating}}  \\ [-2mm]
			\parbox{9.5cm}{
			    \begin{flalign} \label{eq:lpl_1}
                  & \quad j_{lpl} = -2\ a_{n,t}\ i_{0,lpl}\exp\left[-\frac{\alpha_s F}{RT}\left(\Phi_{s,n}-R_{film}I\right)\right] &
				\end{flalign}} \\[-3mm]
			\parbox{9.5cm}{
		         \begin{flalign} \label{eq:lpl_2}
                  & \quad \frac{dc_{Li}}{dt} = -\frac{j_{lpl}}{2F}(1-\beta_{lpl}) &
				\end{flalign}} \\[-1mm]
		
		 			\multicolumn{1}{l}{\textbf{Film thickness}}  \\ [-2mm]	
	         \parbox{9.5cm}{
			\begin{flalign} \label{eq:SEI_3}
			& \quad \frac{dL_{film}}{dt} = \frac{1}{a_{n,t}}\left(\frac{dc_{SEI}}{dt}\frac{M_{SEI}}{\rho_{SEI}} + \frac{dc_{Li}}{dt}\frac{M_{Li}}{\rho_{Li}}\right)=L_{SEI}+L_{Li} &
			\end{flalign}} \\[-2mm]			
				
			\multicolumn{1}{l}{\textbf{Loss of active material}}  \\ [-2mm]
			\parbox{9.5cm}{
			    \begin{flalign} \label{eq:lam_1}
                  & \quad \frac{da_{i,ina}}{dt} = \beta_i'\left(a_i+a_{i,f}-a_{i,ina}\right),\ \ i\in\{p,n\} &\\
                  & \quad a_{i,t} = a_i + a_{i,f} - a_{i,ina} &
				\end{flalign}} \\
			
			\hline\hline
			\end{tabular}}\\ %\vspace{0.5em}
\end{table}

\begin{figure}[!tb]
\centering 
\includegraphics[width=\columnwidth]{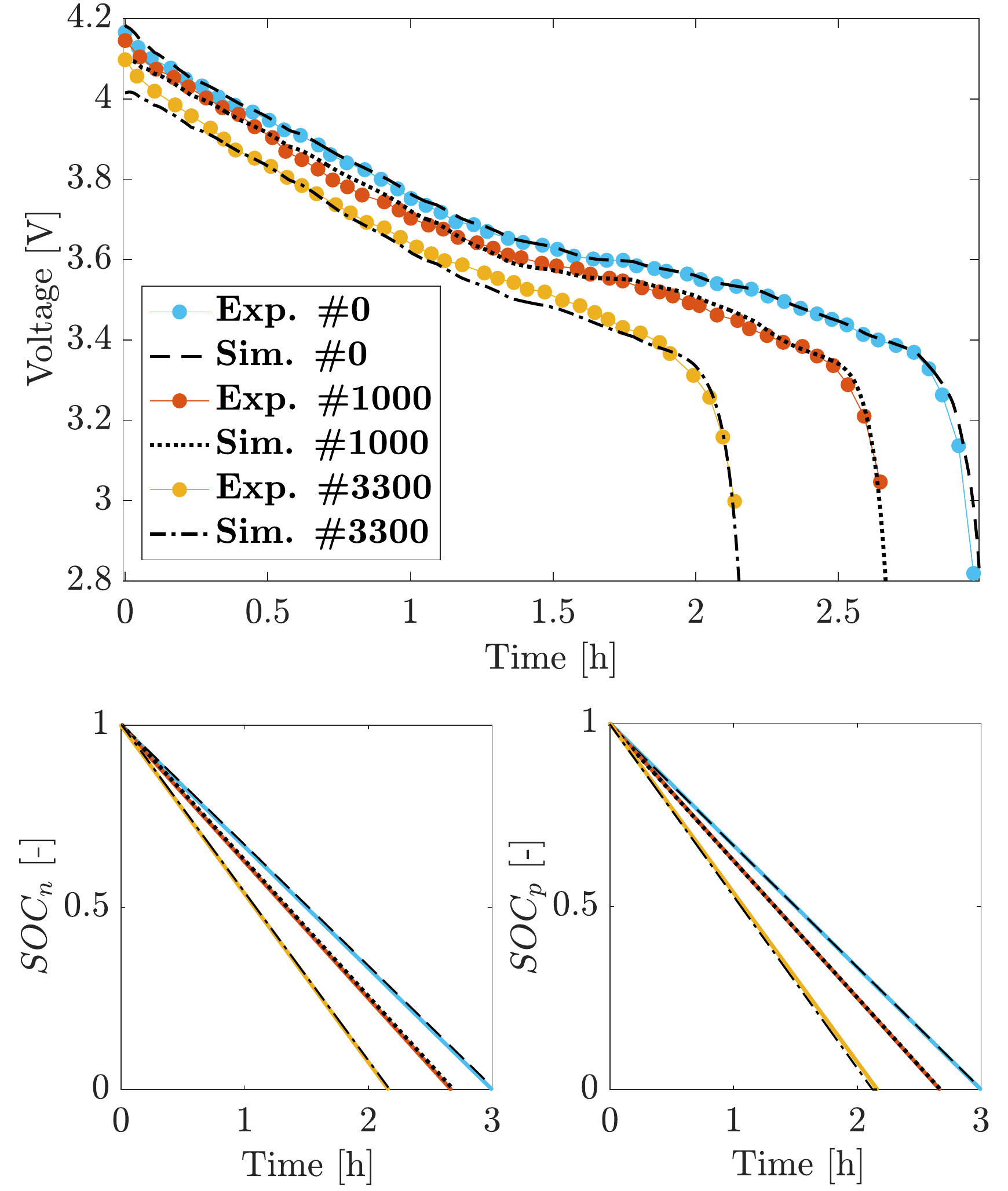}
\caption{\textbf{Performance at different stages of the battery life \cite{pozzato2021modeling}. }  Comparison between C/3 experimental voltage and SOC profile from Coulomb counting with ESPM predicted voltage and electrodes' SOC.}
\label{fig:espmres}	
\end{figure}

% Identification --> Work in progress 
The coupling between mass and charge transport dynamics and aging modes make model parameters identification challenging and,  to tackle this problem,  identification is split between fresh and aged cell conditions.  The optimal parameter vector is identified by minimizing the following cost function:
\begin{equation}
\resizebox{0.75\columnwidth}{!}{$
\begin{split}
J\big(\Theta\big) &=  {w_1 \sqrt{\frac{1}{N} \sum_{j=1}^N\left({V_{exp}(j) - V(\Theta;j)}\right)^2}} \\
&+ w_2 \sqrt{\frac{1}{N} \sum_{j=1}^N(SOC_{CC}(j) - SOC_n(\Theta;j))^2} \\ &+ w_3 \sqrt{\frac{1}{N} \sum_{j=1}^N(SOC_{CC}(j) - SOC_p(\Theta;j))^2}
\end{split}$}
\label{eq:costterms}
\end{equation}
with $\Theta$ the vector collecting parameters to be identified,  $V$ the simulated voltage profile,  $SOC_p$ and $SOC_n$ the simulated SOC at the positive and negative electrodes (computed according to Equation \eqref{goveq:eqsoc}),  $N$ the number of samples,  $j$ the time index,  and $V_{exp}$ and $SOC_{CC}$ the experimental voltage profile and SOC from Coulomb counting,  respectively.  $w_1$ [1/V],  $w_2$ [-],  and $w_3$ [-] are user-defined weights (equal to one) to ensure uniform units of measure. 		\\
The identification process is divided in two phases.  First,  fresh cell data are used to identify the following parameter vector:
\begin{equation}
\Theta_1 = \{A_{cell},R_l,\nu_n, R_p,R_n,D_{s,p}^{ref},D_{s,n}^{ref},\theta_{p,100\%},\theta_{n,100\%}\}
\label{eq:thetaid1}
\end{equation}
where $A_{cell}, R_p$,  and $R_n$ are geometrical parameters,  $\nu_n,D_{s,p}^{ref}$,  and $D_{s,n}^{ref}$ are transport properties\footnote{In \cite{pozzato2021modeling},  solid phase diffusion coefficients are a function of temperature,  and $D_{s,p}^{ref}$ and $D_{s,n}^{ref}$ are the corresponding values at $25^\circ$C. },  $\theta_{p,100\%}$ and $\theta_{n,100\%}$ define the stoichiometric window,  and $R_l$ is the lumped contact resistance.  Secondly,  for aged cells,  the following parameter vector is identified:
\begin{equation}
\Theta_2 = \{L_{SEI}/\kappa_{SEI},\theta_{p,0\%},\theta_{n,100\%}\}
\end{equation}
where $\theta_{n,100\%}$ is reidentified and accounts for the modification of the stoichiometric window due to aging,  and $L_{SEI}/\kappa_{SEI}$ is a lumped parameter associated with SEI formation.  To show the performance of the model,  data from a 12.4Ah NMC/graphite pouch cell discharged at C/3,  at different stages of its life,  are used \cite{yang2017modeling}. The identification is performed assuming that only the negative electrode is aging and the battery is not experiencing LAM.  As shown in Figure \ref{fig:espmres},  performances are satisfactory. 

{In \cite{pozzato2021modeling},  we did not have access to neither the cell open circuit voltage (OCV) nor open circuit potentials and,  for the fresh cell identification,  we adapted the positive electrode open circuit potential with a Newton-like method accounting for the mismatch between simulated and experimental voltage profiles.  This solution could be replaced by the methodology proposed in \cite{lee2019estimation} and \cite{lee2020electrode} if open circuit potentials for the fresh cell and OCV at different stages of life were known. } In these papers,  the battery stoichiometric window and electrode capacities are identified at different stages of the battery life by minimizing the following cost function:
\begin{equation}
J\big(\vartheta\big) =  {\sqrt{\frac{1}{N} \sum_{j=1}^N\left({V_{OCV,exp}(j) - V_{OCV}(\vartheta;j)}\right)^2}}
\label{eq:asid}
\end{equation}
where $\vartheta = \lbrace \theta_{p,100\%}, \theta_{n,100\%},Q_n,Q_p\rbrace$ is the unknown model parameter vector,  and  $V_{OCV,exp}$ and $V_{OCV}$ are the experimental and simulated OCVs,  respectively.  The OCV is the thermodynamic equilibrium potential of the battery and represents the difference in electrical potential between the terminals when no load is present.  In practice,  the experimental OCV -- also known as pseudo-OCV --  is obtained by charging or discharging the battery at low constant currents (e.g.,  C/20). The optimization problem \eqref{eq:asid} is solved while enforcing the following constraints on maximum voltage and stoichiometry at 0\% SOC:
\begin{gather}
V_{OCV}(\vartheta;1) = V_{OCV,exp} (1)\label{eq:ocvmax}\\
 \theta_{n,0\%} = \theta_{n,100\%} - \frac{Q}{Q_n},\ \  \theta_{p,0\%} = \theta_{p,100\%} + \frac{Q}{Q_p}
\end{gather}
with $Q_n$ and $Q_p$ the capacities of negative and positive electrodes, respectively,  and ${Q}$ the cell discharged capacity at a certain point in life.  This formulation of the identification problem and the introduction of constraints \eqref{eq:ocvmax} ensure that simulated and experimental OCVs coincide at the start of the discharge.  Other model parameters included in Equation \eqref{eq:thetaid1},  but not in $\vartheta$,  can be identified considering  additional experimental data (such as the C/3 discharge profiles in Figure \ref{fig:espmres}). 

\section{Machine learning modeling}\label{sec:case}   
\vspace{0.5em}
\begin{mylunchbox}
\textbf{Pros:}
\begin{itemize}
\item Flexible model structure
\item No deep understanding of the physics required
\item Correlations between input and output are extracted directly from data 
\end{itemize}
\end{mylunchbox}

\begin{mylunchbox}
\textbf{Cons:}
\begin{itemize}
\item No physical insight
\item Limited or nonexistent extrapolation capabilities
\item Presence of biases
\item {Large datasets needed (to be fabricated in laboratory)}
\item {Limited or no transferability between chemistries}
\end{itemize}
\end{mylunchbox}

The effectiveness of machine learning modeling has been proven in the assessment of battery SOH -- in terms of impedance,  capacity,  or power  \cite{hu2014method,severson2019data,xing2013ensemble} -- and for RUL prediction\footnote{{Other applications -- e.g.,  material discovery,  manufacturing process modeling/optimization -- exist but are out of the scope of this tutorial paper \cite{lv2022machine,niri2021machine,liu2021feature}.}}.  In this framework,  features are directly extracted from data and used {to assess and/or predict} the battery degradation performance.  The key advantage of this approach is that it removes the need for complex electrochemical models and,  once a model class is chosen,  it relies on experimental/field data and optimization routines (e.g.,  grid search) to tune the model's parameters and select the model's complexity.  {This,  at the same time,  can be a main drawback if data used to train the machine learning model are not representative of the specific battery operation.}

\begin{figure*}[!tb]
\centering 
\includegraphics[width=1\textwidth]{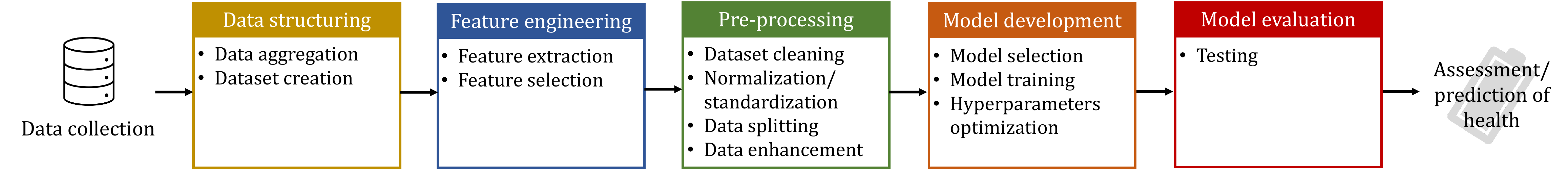}
\caption{{\textbf{Machine learning pipeline. } The pipeline is divided in six steps: data collection,  data structuring,  feature engineering,  pre-processing,  model development,  and model evaluation.}} 
\label{fig:mlpipe}	
\end{figure*}

\begin{figure*}[!tb]
\centering 
\includegraphics[width=0.73\textwidth]{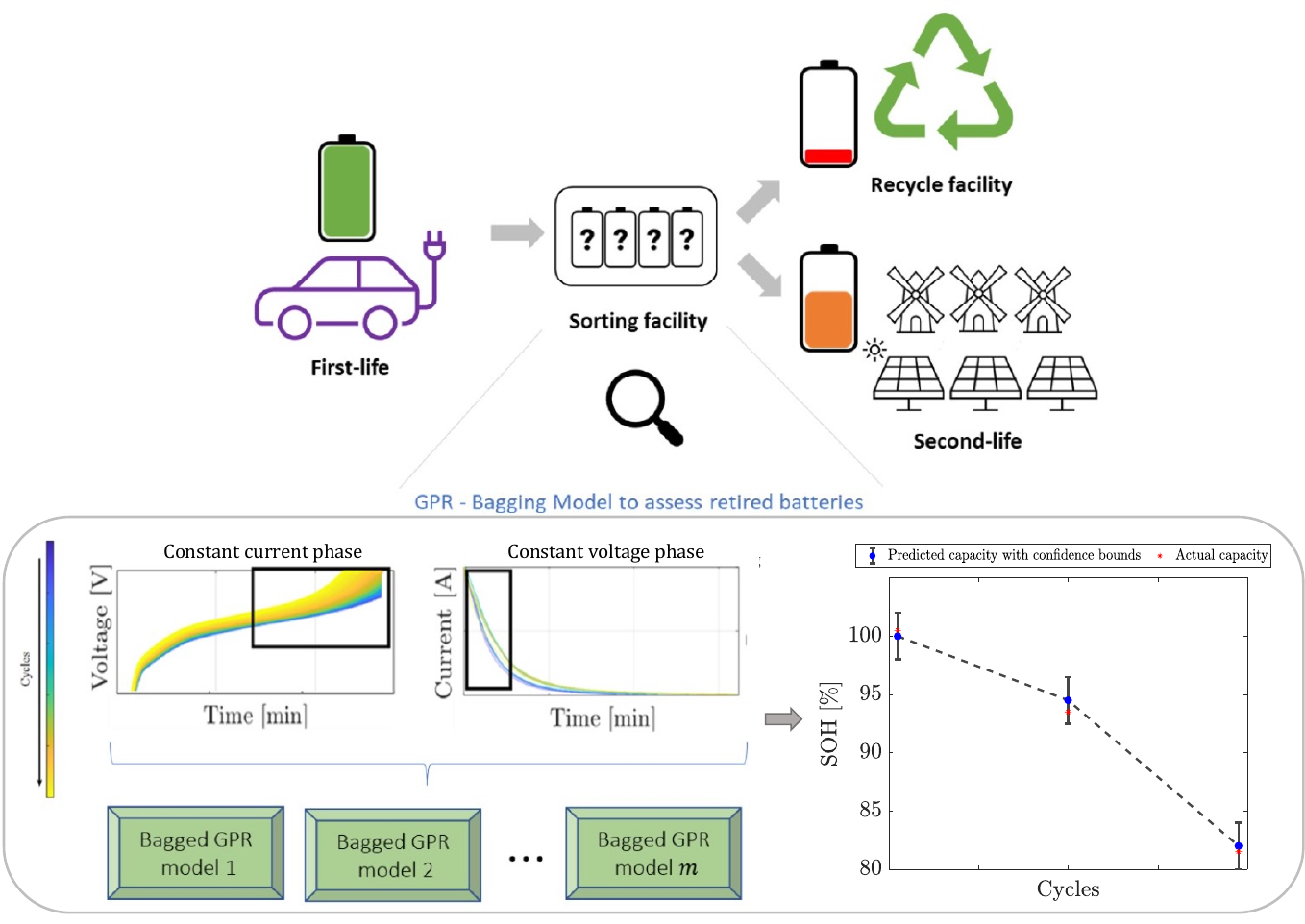}
\caption{{\textbf{SOH estimation framework based on GPR with bagging \cite{takahashi2022second}. } Current (during constant voltage charging) and voltage (during constant current charging) profiles over pre-specified time windows are processed to generate features. These features are used to feed the GPR model with bagging to estimate SOH with uncertainty bounds.}} 
\label{fig:gpraki}	
\end{figure*}

The machine learning model development can be divided in six steps (Figure \ref{fig:mlpipe}): data collection,  data structuring,  feature engineering,  pre-processing,  model development,  and model evaluation. \\

\noindent\textbf{Data collection. } {In the majority of the works,  data is fabricated from laboratory experiments and,  almost never,  is acquired from field operation \cite{aitio2021predicting}.  During this step,  the quality of data is assessed and,  if possible or required by the problem to be solved,  a design of experiments is performed.  The data collection step is crucial because quality and quantity of data directly impact the performance of the machine learning model.  An improper acquisition campaign may lead to biases in the data,  inaccurate data (i.e.,  data are not relevant to the problem to be solved),  missing data,  and data imbalance (i.e.,  some operating conditions are underrepresented). } \\

\noindent\textbf{Data structuring. } {Data is stored and aggregated in structures to create a dataset. } \\

\noindent\textbf{Feature engineering. } Featurization is arguably the backbone of machine learning models development and is composed of feature extraction and/or selection  \cite{khalid2014survey}.   The former is used to reduce the feature space and transform raw data into new attributes while preserving the information of the original dataset.  This can be done with,  for example,  principal component analysis (PCA) \cite{hastie2009elements} or introducing physical understanding to extract physics-informed features from data.  On the other hand,  selection involves the reduction of the number of input variables to reduce the computational time and/or improve the model performance (e.g.,  by removing misleading features).  Feature selection can be performed manually,  for example performing a correlation analysis of the features to select the most relevant ones.  Alternatively,  automatic feature selection algorithms,  e.g.,  the minimum redundancy-maximum relevance (mRMR) algorithm \cite{peng2005feature},  can be used to rank features and identify redundant information to define a subset of attributes to be used as input to the model.\\

\noindent\textbf{Pre-processing. } The dataset is cleaned -- i.e.,  outliers,  null data,  and duplicates are removed -- and scaled through normalization/standardization techniques.  During this step, data is formatted and split into usable data structures.  Data enhancement (such as interpolation) could also be included. \\

\noindent\textbf{Model development. } Depending on the chosen model class,  machine learning models have different structures.  In this tutorial,  we define the generic input/output relationship as follows:
\begin{equation}
y = \mathcal{F}(\Psi)
\end{equation}
where $y$ is the output,  and $\mathcal{F}$ a generic function linking features $\Psi$ to $y$.  At this stage,  the dataset is divided into training,  validation (to fine tune the model's hyperparameters),  and testing.  During training,  hyperparameters optimization algorithms such as grid search and Bayesian optimization \cite{feurer2019hyperparameter} are used to tune the model and choose its order inside the chosen class (e.g.,  number of hidden layers for neural networks). \\

\noindent\textbf{Model evaluation. } Performances of the machine learning model are quantified by means of metrics like root mean squared error (RMSE),  root mean squared percentage error (RMSPE),  and maximum absolute percentage error (MAPE) \cite{murphy2012machine}.

\subsection{Case study: Battery health estimation for second-life applications}\label{cs2}

After retirement,  batteries are sent to warehouses where they are stored waiting to be screened.  A key issue with these batteries is that health history is unknown and algorithms are needed to assess the level of degradation to decide whether the cell can be safely reused in a second-life application (e.g.,  backup power,  residential storage,  electric vehicle (EV) charging,  and utility scale storage \cite{moy2021design}, \cite{moy2023}) or recycled to recover materials.  According to \cite{engel2019second},  second-life batteries could surpass 200GWh by 2030,  with the potential to meet half of the global demand for utility-scale energy storage in that year. 

\vspace{0.35em}\noindent\textit{A.1 Offline solution}\vspace{0.2em}

Different approaches have been explored to solve the problem of health assessment for batteries' first life.  In \cite{mansouri2017remaining} and \cite{nuhic2013health},   support vector regression and random forest are used for SOH assessment and RUL prediction.  In \cite{yang2017neural} and \cite{roman2021machine},  neural networks are used to find relationships between degradation indicators and battery health.  Gaussian process regression (GPR) has become an increasingly popular method because it is characterized by interpretability of results and quantification of prediction uncertainty \cite{hu2014method,richardson2017gaussian,liu2013prognostics,huonori2020battery}.  Different applications of GPR to the battery health assessment problem exist.  {In \cite{yang2018novel} and \cite{yu2018state},  under the assumption that batteries undergo similar use cases in both early and late cycling (inappropriate for second-life battery applications),  training is performed on some early cycles and health is estimated thereafter.   In \cite{richardson2017gaussian},  models are trained on a subset of cells and health is assessed on different test cells operated in similar conditions.  }

{The nonlinear nature of battery degradation dynamics poses a challenge in health estimation for second-life applications. } Scalability of machine learning methods is also an issue and must be addressed to allow the inflow of battery data.  {To tackle these issues,  in \cite{takahashi2022second},  we use an ensemble learning model in the form of bagging embedded with GPR.} This approach allows to estimate cell capacity to assess the residual health of retired batteries,  and uses features from voltage and current in a limited window of the charge profile.  In the proposed approach,  time is not explicitly used because it is impractical when the battery experiences incomplete charge or discharge.  The algorithm presented in \cite{takahashi2022second},  summarized in Figure \ref{fig:gpraki},  assesses the feasibility of retired batteries for second-life applications,  providing a  decision on recycling and repurposing.  The algorithm is designed to operate in a battery recycling,  repurposing,  or sorting facility where retired batteries are sent for screening.  

\begin{figure}[!tb]
\centering 
\includegraphics[width=0.8\columnwidth]{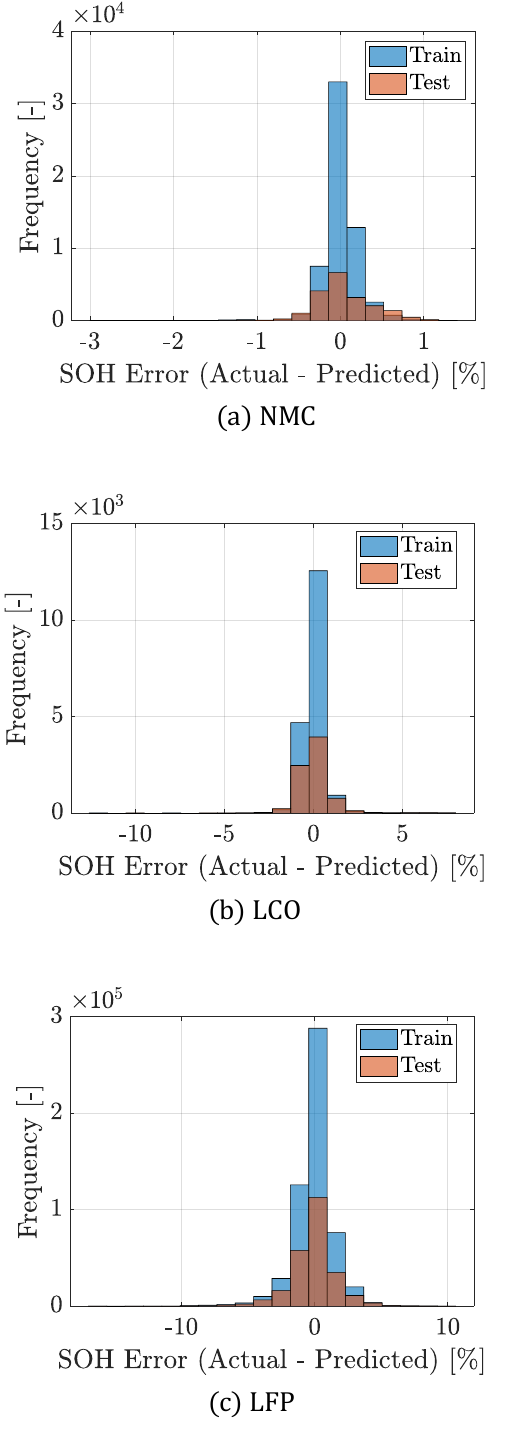}
\caption{{\textbf{SOH estimation performance \cite{takahashi2022second}.}  Distribution of SOH error for training and testing datapoints for (a) NMC \cite{birkl2017oxford}, (b) LCO \cite{bole2014adaptation}, (c) LFP cells \cite{severson2019data,attia2020closed}.}} 
\label{fig:gprresults}	
\end{figure}

{Features are extracted from voltage (during constant current charging) and current (during constant voltage charging) and used to train the algorithm with bagging,  which is used to estimate the state of health defined in terms of capacity loss:  }
\begin{equation}
SOH = \frac{Q}{Q_{nom}}\cdot 100\%
\end{equation}
where $Q$ is the capacity at a certain point in time,  and $Q_{nom}$ is the nominal fresh cell capacity.
The algorithm is trained and tested over three different chemistries,  namely 0.74Ah NMC pouch cells \cite{birkl2017oxford},  2.1Ah LCO cylindrical cells \cite{bole2014adaptation},  and 1.1Ah LFP cylindrical cells \cite{severson2019data,attia2020closed}.  As shown in  Figure \ref{fig:gprresults}	,  the proposed approach is robust and allows for the estimation of battery SOH with a root mean square percent error of 0.25\%,  1.28\%,  and 1.48\% for NMC,  LCO,  and LFP,  respectively.  {Specifically,  bagging allows to reduce the estimates' variance because,  instead of relying on a single model,  it creates several models working in parallel improving accuracy and reducing overfitting.}
{These promising results make the proposed framework a suitable candidate for the screening of retired batteries. }

\vspace{0.35em}\noindent\textit{A.2 Online solution}\vspace{0.2em}

The approach  in \cite{takahashi2022second} is proposed for offline health estimation and cell screening.  However,  the problem of estimating health during second-life operation {is still a current challenge that hinders the deployment of retired batteries in grid applications}.  To solve this challenge,  it is crucial to develop strategies based on field data,  accounting for history,  and adapted over time.  {Current work in our laboratory is focusing on machine learning for health estimation in retired LIBs  \cite{cuirelyion} and on creating a dataset collecting laboratory data of batteries cycled with grid application profiles \cite{kevindib}.  In \cite{cuirelyion},  the approach was first developed offline and is now being explored for online implementation and use.  }

\begin{figure}[!tb]
\centering 
\includegraphics[width=\columnwidth]{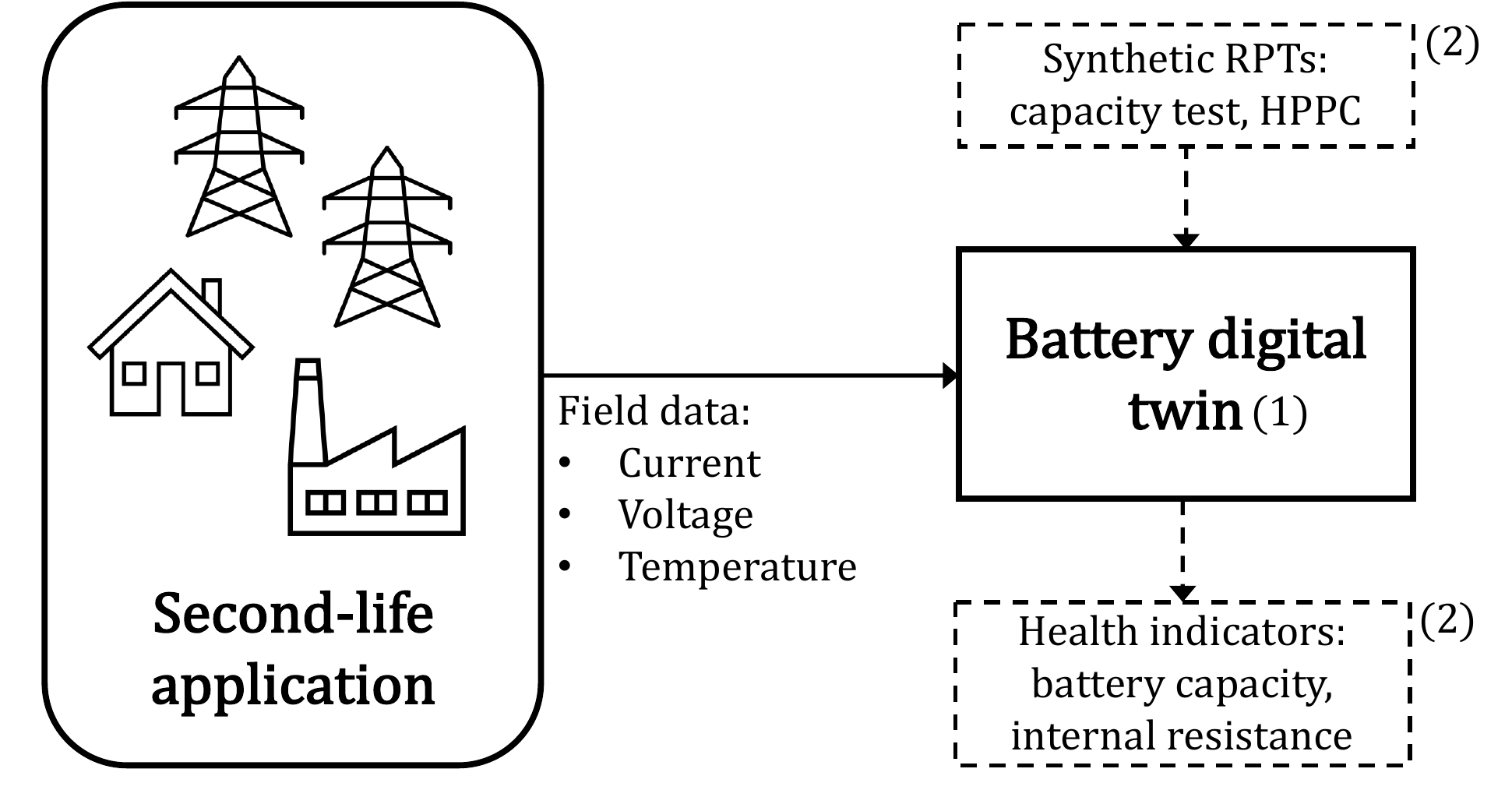}
\caption{\textbf{Online SOH estimation for second-life applications.} A digital twin-based solution for online battery health assessment in second-life applications.} 
\label{fig:slonline}	
\end{figure}

We recall that,  in laboratory settings,  battery health is generally assessed with reference performance tests (RPTs) such as capacity tests and hybrid pulse power characterization (HPPC) \cite{christophersen2015battery}.  Capacity tests consist of constant current charge and discharge protocols performed at low current (e.g.,  C/20)  and used to assess the battery capacity.  The HPPC is a test profile incorporating both discharge and regeneration pulses at various SOC levels, which allows for the assessment of the power capability of a cell.  RPTs are effective ways to assess the battery aging status,  however,  they can not be performed on-board because of time constraints: the duration of a C/20 charge or discharge capacity test is 20 hours.   Inspired by \cite{heinrich2022virtual},  we propose to solve this issue developing a battery cell digital twin.  Starting from field data,  i.e.,  current,  temperature,  and voltage,  data-driven models such as long short-term memory (LSTM) networks or autoregressive–moving-average (ARMAX) models can be used to learn the input/output behavior and build a battery digital twin (Figure \ref{fig:slonline}(1)).  Instead of performing RPTs in a laboratory,  we use the digital twin to run synthetic RPTs (Figure \ref{fig:slonline}(2)) and extract information on health, e.g.,  in the form of charged/discharged capacity from capacity tests.  This approach,  yet to be proved,  could be an effective tool to estimate the battery health on-board and when history is lacking.

\section{Hybrid modeling}
\vspace{0.5em}
\begin{mylunchbox}
\textbf{Pros:}
\begin{itemize}
\item Extrapolation capabilities
\item Retaining of the physics
\item {Transferability across chemistries}
\item {Machine learning adaptation and/or compensation of unmodeled physics}
\end{itemize}
\end{mylunchbox}

\begin{mylunchbox}
\textbf{Cons:}
\begin{itemize}
\item No unique formulation and implementation
\item {Customized for the application}
\end{itemize}
\end{mylunchbox}

Hybrid modeling is a promising,  still underexplored paradigm that entails the blending of physics-based techniques with machine learning.  Compared to Section \ref{sec:pb} and \ref{sec:case},  where a structured workflow {or a well-defined set of equations} are provided,   research on hybrid models is in its early stages and a variety of architectures can be used.  In \cite{aykol2021perspective},  a comprehensive list of system level architectures to create hybrid models is provided.  While no implementation is given,  this reference is a good starting point to define architectures for hybrid models to be customized for the target application.   An example of hybrid battery model is found in \cite{park2017hybrid}  where the integration of SPM with neural networks is proposed in a simulation environment and used to cope with model uncertainties.  Here,  the true battery behavior is learned  from the DFN model, which on the other hand, suffers from lack of predictability over certain ranges of SOC, C-rate and temperature operation,   \cite{arunachalam2019full}. 
In \cite{chu2020stochastic},  an aging model is used to generate synthetic  capacity loss data to complement an experimental dataset and develop a stochastic health prediction model.  A similar approach is used in \cite{jia2021data} and \cite{jia2022precise},  where the physics-based model is used to generate synthetic data to feed a machine learning model predicting the safety risk in LIBs. 

According to \cite{aykol2021perspective},  hybridization could tap into the advantages of both physics-based and machine learning models,  retaining physics while exploiting machine learning to describe complex and poorly understood physical phenomena (such as battery degradation) that would be hard to characterize by means of first principles.  

\begin{figure*}[!tb]
\centering 
\includegraphics[width=1\textwidth]{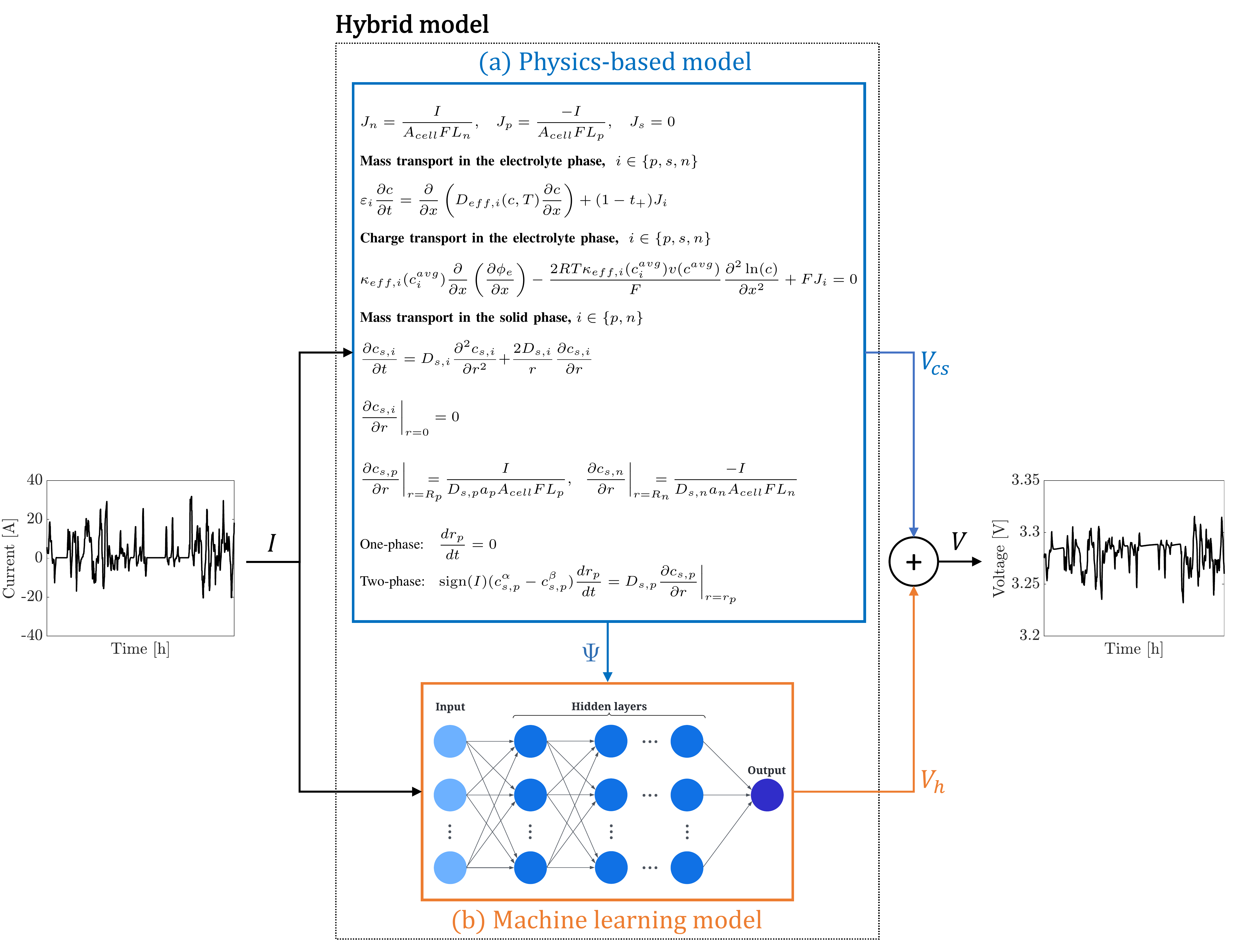}
\caption{\textbf{Hybrid model architecture \cite{pozzatopreprint}. } The physics-based model is used to describe the underlying electrochemical dynamics (a) and the machine learning model describes hysteresis (b).}
\label{fig:hybrid}	
\end{figure*}

\subsection{Case study: Hysteresis modeling in LiFePO$_4$/graphite batteries}\label{cs3}
LFP batteries use graphite and $\mathrm{LiFePO}_4$  as negative and positive electrodes active materials,  respectively.  The positive electrode is characterized by a two-phase  transition region with the coexistence of a lithium-rich phase $\mathrm{Li_{\beta}FePO_4}$ ($\beta\simeq 1$) and a lithium-poor phase $\mathrm{Li_{\alpha}FePO_4}$ ($\alpha \simeq 0$) \cite{love2013,yamada2005}.  In the positive electrode,  the presence of a two-phase transition generates a flat OCV profile,  which makes the estimation of SOC challenging because of the lack of system's states observability \cite{he2020modeling}.  According to \cite{ovejas2019effects},  hysteresis results from mechanical stress and thermodynamic effects.  Mechanical hysteresis is related to the different lattice constants of lithiated and delithiated phases causing mechanical stress at the phase boundary.  Thermodynamic effects are associated with heterogeneous lithium insertion rates in the electrode.  

In LFP batteries,  the positive electrode is the principal source of OCV hysteresis which,  in \cite{srinivasan2006existence},  is described as lithium-poor and lithium-rich phases coexisting in a core-shell structure.  During discharge,  core and shell are characterized by a lithium-poor and a lithium-rich phase,  respectively,  and the opposite happens during charge.  Notably,  in \cite{dreyer2010thermodynamic},  the thermodynamic origin of hysteresis is associated with the presence of different lithium insertion rates in the positive electrode active material.  Specifically,  non-uniform insertion rates lead to heterogeneous lithium concentrations,  non-uniform potentials,  and hysteresis.  

\begin{figure}[!tb]
\centering 
\includegraphics[width=0.85\columnwidth]{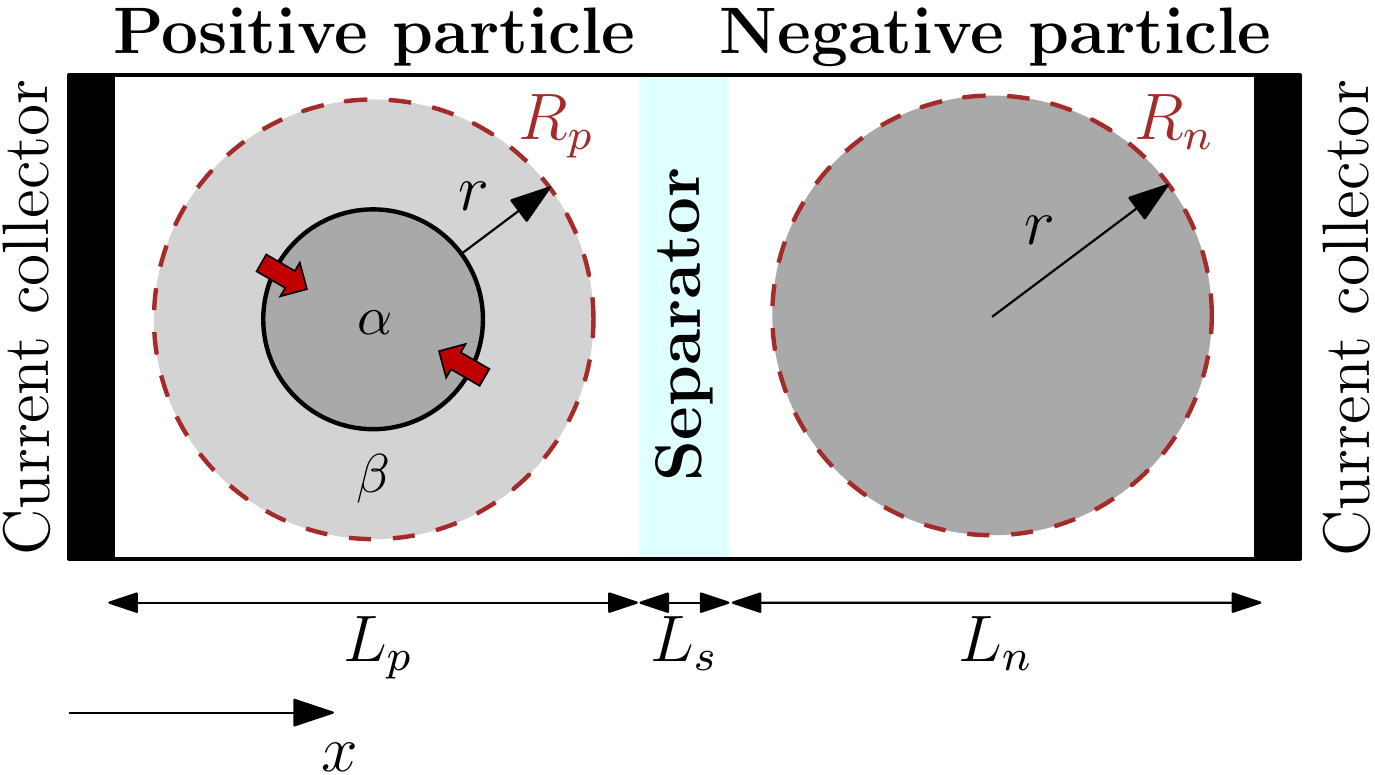}
\caption{{\textbf{Battery schematic for core-shell ESPM. } In the core-shell ESPM,  positive and negative electrodes are modeled as single particles divided by a separator and immersed in the electrolyte.} Considering a discharge condition,  a core-shell modeling paradigm is used to describe the phase transition from $\alpha$ (core) to $\beta$ (shell) in the positive electrode.  During charge,  the opposite happens,  with the core in $\beta$-phase being replaced by the $\alpha$ shell.  }
\label{fig:espm_cs}	
\end{figure}

The flat OCV and hysteresis challenge the current modeling of LIBs.  To solve this problem,  in \cite{pozzatopreprint},  we develop a hybrid model  leveraging the average core-shell ESPM model \cite{pozzato2022core,taka2022},  to track the positive electrode phase transition,  and machine learning,  to describe voltage hysteresis.  As shown in Figure \ref{fig:hybrid},  the proposed hybrid framework blends physics with machine learning to create an accurate battery model capturing the system's hysteresis.  In Figure \ref{fig:hybrid}(a),  the physics-based model is developed based on Equations \eqref{eq:mb1},  \eqref{eq:mb2},  and \eqref{eq:cb1},  with the addition of the following mass balance describing the two phase transition within the positive electrode (shown in Figure \ref{fig:espm_cs}):
\begin{equation}
\mathrm{sign}(I)(c_{s,p}^\alpha-c_{s,p}^\beta)\frac{d r_p}{d t} = D_{s,p}\frac{\partial c_{s,p}}{\partial r}\bigg|_{r = r_p}
\end{equation}
with $r_p$  the moving boundary,  i.e.,   the distance between the center of the positive particle and the interface between $\alpha$- and $\beta$-phase,  $c_{s,p}^\alpha$ and $c_{s,p}^\beta$ the concentrations in $\alpha$- and $\beta$-phase,  respectively,  and $\mathrm{sign}(I)$ accounting for the fact that,  during charge,  $\beta$-phase transitions to $\alpha$-phase and that the opposite happens during discharge.  \\
In \ref{fig:hybrid}(b),  the machine learning model compensates the output of the physics based model by reconstructing the hysteretic behavior from the experimental current profile $I$ and the simulated vector $\Psi$,  both used as input features.  Three classes of machine learning models are used to describe the battery hysteresis $V_h$,  namely feedforward neural networks,  regression trees,  and random forests.  \\

\begin{figure*}[!tb]
\centering 
\includegraphics[width=1\textwidth]{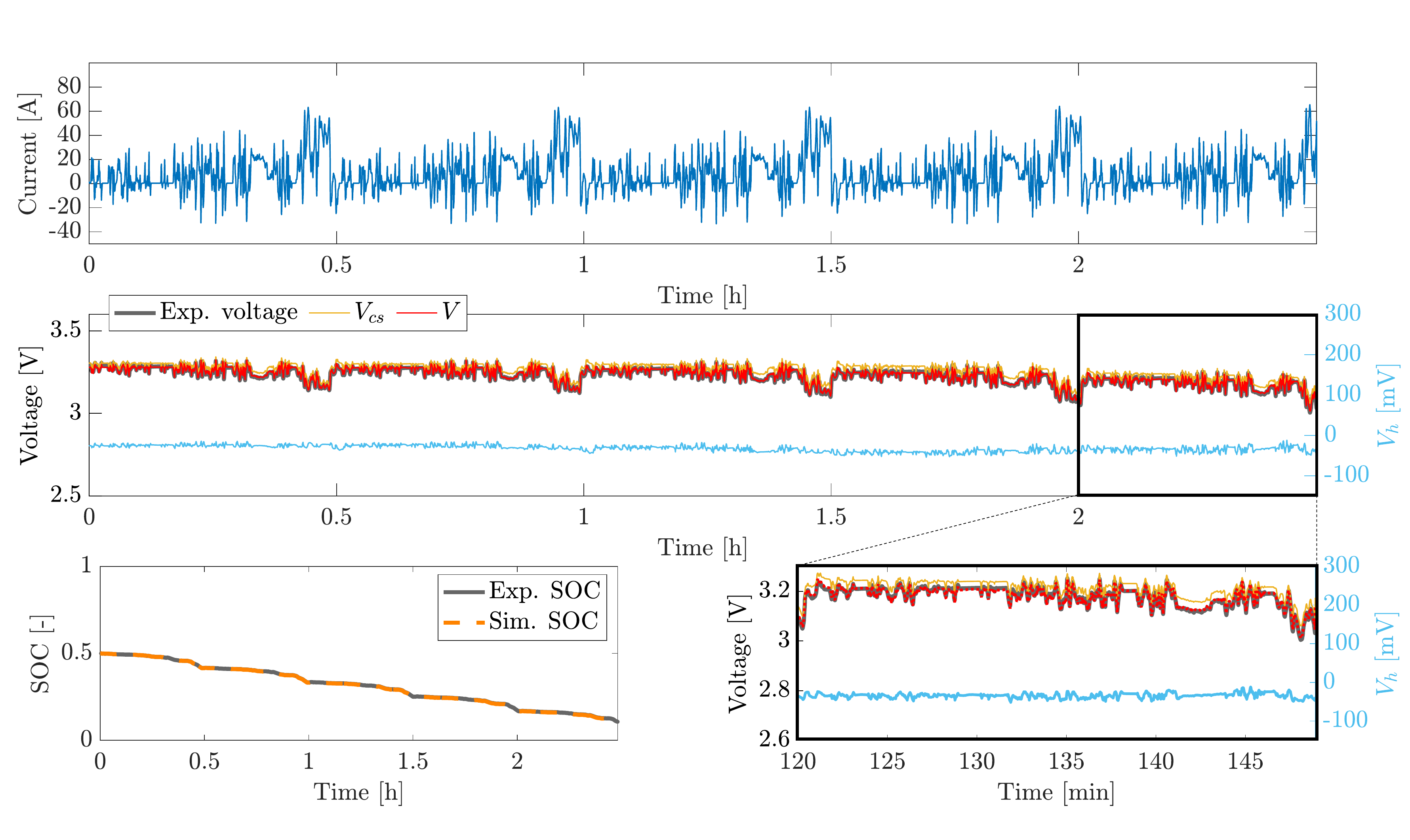}
\caption{\textbf{Hybrid model validation  \cite{pozzatopreprint}. } Application of the hybrid modeling framework to the testing driving cycle data.  In this scenario,  a random forest is used to model hysteresis. } 
\label{fig:eghybrid}	
\end{figure*}

The output of the hybrid model is the battery terminal voltage given by the summation of the physics-based ($V_{cs}$) and machine learning ($V_h$) contributions:
\begin{equation}
\resizebox{\columnwidth}{!}{$
\begin{split}
V & = V_{cs} + V_h = \\
& = \underbrace{(U_p^{ch} + U_p^{dis})/2 - U_n + \eta_p - \eta_n +\Delta\Phi_e - I\cdot R_l(SOC,I)}_{V_{cs}} + V_h = \\
& = \underbrace{U_{p}^{avg} - U_n + \eta_p - \eta_n +\Delta\Phi_e - I\cdot R_l(SOC,I)}_{V_{cs}} + V_h =\\
& = \underbrace{V_{OCV}^{avg} + \eta_p - \eta_n +\Delta\Phi_e - I\cdot R_l(SOC,I)}_{V_{cs}} + V_h 
\end{split}$}
\label{eq:hybridoutput}
\end{equation}
where the $V_{cs}$ collects the information from the  physics-based model  and $V_h$ models hysteresis.  $V_{cs}$ is a function of the positive and negative electrode overpotentials ($\eta_p$ and $\eta_n$),  the electrolyte overpotential ($\Delta\Phi_e$),  the negative electrode open circuit potential ($U_n$),  and the positive electrode charge ($U_p^{ch}$) and discharge ($U_p^{dis}$) open circuit potentials \cite{plett2015battery}.  $U_p^{ch}$ and $U_p^{dis}$ are used to compute the average positive particle open circuit potential.  The Ohmic loss term ($I\cdot R_{l}$) is accounting for the battery high frequency resistance (lumping both contact resistance and electro-migration in the electrolyte phase) and is a function of the input current and SOC.  

A key characteristic of the machine learning component is that hysteresis is learnt from both simulated features (from the physics-based model) and experimental data (the current profile).  Machine learning models are trained and validated over 19 and 15 hours of EV real-driving profiles acquired testing a 49Ah LiFePO$_4$/graphite pouch cell.  To assess the hybrid model performance,  the RMSE between the experimental voltage and output of the hybrid model is computed.  Compared to a purely physics-based model,  the hybrid framework leads to a visible improvement of the voltage RMSEs: 95\% for training datasets,  and between 83\% and 47\% for testing.  The random forest is the modeling strategy providing the best performance over both training and testing datasets.  This is reasonable because the random forest is an ensemble of regression trees where the output is computed as the average of the predictions from each tree.   Figure \ref{fig:eghybrid}	 shows a comparison between experimental data and simulation results with and without the machine learning hysteresis model.  As shown in the zoomed portion,  the machine learning model compensates for the bias introduced by hysteresis,  reducing the modeling error.

\section{Conclusions and perspectives}\label{sec:concl}
In this tutorial paper,  we analyzed three different strategies to tackle problems related to battery modeling and estimation.  Electrochemical models are effective approaches to describe the physics involved in the operation of LIBs,  however,  parameter identifiability might be a limiting factor in particular when degradation modes are integrated.  Hence,  it is key to carefully design experiments for robust calibration of the model.  The principal advantage of machine learning is that it can uncover hidden correlations,  recognize high-dimensional patterns,  or even describe poorly understood phenomena directly from data and with a flexible modeling structure.  In this context,  the quality and quantity of data play a crucial role,  and the limited or nonexistent extrapolation capability limit their application.  Finally,  hybrid models blend physics with machine learning,  tapping into the advantages of both strategies.

The selection of the most appropriate modeling approach is a function of the problem and the quantity and quality of data at hand.  Nevertheless,  all these approaches require experiments or field data in order to calibrate parameters and hyperparameters.  Therefore,  it is crucial to use a \myquote{data-centric} approach and design informative experiments which are representative for the target application.  In this context,  results from a recent work in our laboratory  motivate the need to enhance battery laboratory testing methods by deliberately incorporating effects of varying temperature into the design of experiments to generate meaningful datasets that can be used to develop accurate models and on-field performance estimation algorithms \cite{allamnc}. 

\section*{Acknowledgment}
{We would like to mention that this tutorial paper summarizes a collaborative effort of the Stanford Energy Control Lab (\href{https://onorilab.stanford.edu/}{SECL})  members and alumnae/alumni.  The research presented in this tutorial is partially supported by the Bits and Watts Initiative and StorageX Initiative within the Precourt Institute for Energy at Stanford University,  LG Energy Solutions,  and Volkswagen.}

{The authors would also like to thank Prof.  Anna Stefanopoulou (University of Michigan) for the insightful discussions on electrochemical models identification and identifiability analysis.  }

\begin{nomenclature}
\small{
\entry{$brugg$}{Bruggeman coefficient, $[\mathrm{-}]$}
\entry{$j$}{Time index,  $[\mathrm{-}]$}
\entry{$N$}{Number of samples,  $[\mathrm{-}]$}
\entry{$SOC_i$}{Simulated state of charge, $[\mathrm{-}]$}
\entry{$SOC_{CC}$}{State of charge from Coulomb counting, $[\mathrm{-}]$}
\entry{$SOH$}{State of health, $[\mathrm{-}]$}
\entry{$t_+$}{Transference number, $[\mathrm{-}]$}
\entry{$v$}{Thermodynamic factor, $[\mathrm{-}]$}
\entry{$\alpha_a,\alpha_c$}{Anodic and cathodic charge transfer coefficient, $[\mathrm{-}]$}
\entry{$\alpha_s$}{Side reactions charge transfer coefficient, $[\mathrm{-}]$}
%\entry{$\nu_{i,filler}$}{Filler active volume fraction [$i \in\hat{\mathcal{M}}$], $[\mathrm{-}]$}
\entry{$\beta_i'$}{Inactive area evolution coefficient, $[\mathrm{-}]$}
\entry{$\beta_{lpl}$}{Fraction of lithium plating converted into SEI, $[\mathrm{-}]$}
%\entry{$k_i^{'}$}{Fracture evolution coefficient, $[\mathrm{-}]$}
\entry{$\varepsilon_i$}{Porosity, $[\mathrm{-}]$}
\entry{$\varepsilon_{i,0}$}{Initial porosity, $[\mathrm{-}]$}
\entry{$\theta_i$}{Bulk normalized lithium concentration, $[\mathrm{-}]$}
\entry{$\theta_{i,0\%}$}{Reference stoichiometry ratio at 0\% $SOC$ $[\mathrm{-}]$}
\entry{$\theta_{i,100\%}$}{Reference stoichiometry ratio at 100\% $SOC$, $[\mathrm{-}]$}
\entry{$\nu_{i}$}{Solid phase active volume fraction, $[\mathrm{-}]$}
\entry{$t$}{Time, $[\mathrm{s}]$}
\entry{$L_{film}$}{Thickness of the surface film, $[\mathrm{m}]$}
\entry{$L_i$}{Region thickness, $[\mathrm{m}]$}
\entry{$L_{Li}$}{Fraction of $L_{film}$ from lithium plating, $[\mathrm{m}]$}
\entry{$L_{SEI}$}{Fraction of $L_{film}$ from SEI growth, $[\mathrm{m}]$}
\entry{$r$}{Radial coordinate, $[\mathrm{m}]$}
\entry{$r_p$}{Moving boundary, $[\mathrm{m}]$}
\entry{$R_i$}{Particle radius, $[\mathrm{m}]$}
\entry{$x$}{Cartesian coordinate, $[\mathrm{m}]$}
\entry{$A_{cell}$}{Cell cross section area, $[\mathrm{m}^2]$}
\entry{$a_{i}$}{Specific surface area, $[\mathrm{m}^2]$}
\entry{$a_{i,f}$}{Specific fracture surface area, $[\mathrm{m}^2/\mathrm{m}^3]$}
\entry{$a_{i,ina}$}{Specific inactive surface area, $[\mathrm{m}^2/\mathrm{m}^3]$}
\entry{$a_{i,t}$}{Total specific surface area, $[\mathrm{m}^2/\mathrm{m}^3]$}
\entry{$k_{f}$}{SEI side reaction kinetic constant, $[\mathrm{m/s}]$}
\entry{$D_i$}{Electrolyte phase diffusion coefficient, $[\mathrm{m^2/s}]$}
\entry{$D_{eff,i}$}{Effective electrolyte phase diffusion coefficient, $[\mathrm{m^2/s}]$}
\entry{$D_{s,i}$}{Solid phase diffusion coefficient, $[\mathrm{m^2/s}]$}
%\entry{$D_{s,i}^{ref}$}{Reference solid phase diffusion coefficient [$i \in\hat{\mathcal{M}}$], $[\mathrm{m^2/s}]$}
%\entry{$D_{solv}$}{Solvent diffusion coefficient in SEI layer, $[\mathrm{m^2/s}]$}
\entry{$c$}{Electrolyte concentration, $[\mathrm{mol/m^3}]$}
\entry{$c_{Li}$}{Plated lithium concentration, $[\mathrm{mol/m^3}]$}
%\entry{$c^{avg}$}{Average electrolyte concentration, $[\mathrm{mol/m^3}]$}
\entry{$c_{s,i}$}{Solid phase concentration, $[\mathrm{mol/m^3}]$}
\entry{$c_{s,p}^{\alpha},c_{s,p}^{\beta}$}{Positive particle solid phase concentration in $\alpha$ and $\beta$ phases, $[\mathrm{mol/m^3}]$}
%\entry{$c_{s,i}^{surf}$}{Surface solid phase concentration [$i \in\hat{\mathcal{M}}$], $[\mathrm{mol/m^3}]$}
%\entry{$c_{s,i}^{max}$}{Maximum solid phase concentration [$i \in\hat{\mathcal{M}}$], $[\mathrm{mol/m^3}]$}
\entry{$c_{SEI}$}{SEI concentration, $[\mathrm{mol/m^3}]$}
\entry{$c_{solv}$}{Solvent concentration, $[\mathrm{mol/m^3}]$}
%\entry{$c_{eff,solv}^{bulk}$}{Effective solvent bulk concentration [function of the film porosity \cite{prada2013simplified}], $[\mathrm{mol/m^3}]$}
\entry{$k_{i}$}{Reaction rate, $[\mathrm{m^{2.5}/(mol^{0.5}\cdot s)}]$}
\entry{$J_{i}$}{Pore wall flux, $[\mathrm{mol/(m^3\cdot s)}]$}
\entry{$I$}{Applied current, $[\mathrm{A}]$}
\entry{$Q$}{Capacity at a certain point in the battery life,  $[\mathrm{Ah}]$}
\entry{$Q_i$}{Electrode capacity,  $[\mathrm{Ah}]$}
\entry{$Q_{nom}$}{Nominal fresh cell capacity,  $[\mathrm{Ah}]$}
\entry{$i_{0,i}$}{Exchange current, $[\mathrm{A/m^2}]$}
\entry{$i_{0,lpl}$}{Lithium deposition exchange current, $[\mathrm{A/m^2}]$}
%\entry{$j_{int}$}{Intercalation current density, $[\mathrm{A/m^3}]$}
\entry{$j_{lpl},j_{SEI}$}{Side reactions current densities, $[\mathrm{A/m^3}]$}
\entry{$\kappa$}{Electrolyte phase conductivity,  $[\mathrm{S/m}]$}
\entry{$\kappa_{eff,i}$}{Effective electrolyte phase conductivity, $[\mathrm{S/m}]$}
\entry{$\kappa_{SEI}$}{SEI ionic conductivity, $[\mathrm{S/m}]$}
\entry{$R_{el}$}{Electrolyte resistance, $[\Omega]$}
\entry{$R_{film}$}{Film resistance, $[\Omega]$}
\entry{$R_{l}$}{Lumped contact resistance, $[\Omega]$}
%\entry{$\Phi_{s,n}$}{Solid phase potential at the anode, $[\mathrm{V}]$}
%\entry{$\Phi_{e,n}$}{Liquid phase potential at the anode, $[\mathrm{V}]$}
\entry{$U_i$}{Open circuit potential,  $[\mathrm{V}]$}
\entry{$V$}{Cell voltage, $[\mathrm{V}]$}
\entry{$V_{cs}$}{Cell voltage from physics-based core-shell model, $[\mathrm{V}]$}
\entry{$V_h$}{Voltage hysteresis, $[\mathrm{V}]$}
\entry{$V_{OCV}$}{Cell OCV, $[\mathrm{V}]$}
\entry{$\eta_i$}{Overpotential, $[\mathrm{V}]$}
\entry{$\phi_{e}$}{Liquid phase potential, $[\mathrm{V}]$}
\entry{$\Phi_{s,n}$}{Solid phase potential at the negative electrode defined as ($U_n+\eta_n+R_{film}I$), $[\mathrm{V}]$}
\entry{$\Delta\Phi_e$}{Diffusion overpotential, $[\mathrm{V}]$}
\entry{$T$}{Battery cell temperature, $[\mathrm{K}]$}
%\entry{$T_{ref}$}{Battery cell reference temperature, $[\mathrm{K}]$}
\entry{$R$}{Universal gas constant, $[\mathrm{J/(mol\cdot K)}]$}
%\entry{$E_{a,D_s}^i$}{Activation energy of solid diffusivity [$i \in\hat{\mathcal{M}}$], $[\mathrm{J/mol}]$}
\entry{$F$}{Faraday constant, $[\mathrm{C/mol}]$}
\entry{$M_{Li}$}{Lithium molar mass, $[\mathrm{kg/mol}]$}
\entry{$M_{SEI}$}{SEI molar mass, $[\mathrm{kg/mol}]$}
\entry{$\rho_{Li}$}{Lithium density, $[\mathrm{kg/m^3}]$}
\entry{$\rho_{SEI}$}{SEI density, $[\mathrm{kg/m^3}]$}}
\end{nomenclature}

\begin{notation}
\small{
\entry{$f,\mathcal{F}$}{Generic functions}
\entry{$\vartheta,\Theta$}{Model parameter vectors}
\entry{$\Psi$}{Feature vector}
\entry{$i$}{Index indicating the domain: $p$, $n$, or $s$}
\entry{$n$}{Negative electrode}
\entry{$p$}{Positive electrode}
\entry{$s$}{Separator}
\entry{$y$}{Generic output of a machine learning model}
\entry{$avg$}{Average quantity}
\entry{$ch$}{Charge}
\entry{$dis$}{Discharge}
\entry{$exp$}{Experimental data}
\entry{$ref$}{Reference temperature}
\entry{$surf$}{Particle surface}}
\end{notation}

\bibliographystyle{IEEEtran}
\bibliography{biblio} 

\end{document}